\documentstyle[wstwocl,epsfig]{article}
\begin{document}

\bibliographystyle{unsrt}    % for BibTeX - sorted numerical labels

\newcommand{\st}{\scriptstyle}
\newcommand{\sst}{\scriptscriptstyle}
\newcommand{\mco}{\multicolumn}
\newcommand{\epp}{\epsilon^{\prime}}
\newcommand{\vep}{\varepsilon}
\newcommand{\ra}{\rightarrow}
\newcommand{\ppg}{\pi^+\pi^-\gamma}
\newcommand{\vp}{{\bf p}}
\newcommand{\ko}{K^0}
\newcommand{\kb}{\bar{K^0}}
\newcommand{\al}{\alpha}
\newcommand{\ab}{\bar{\alpha}}
\def\be{\begin{equation}}
\def\ee{\end{equation}}
\def\bea{\begin{eqnarray}}
\def\eea{\end{eqnarray}}
\def\CPbar{\hbox{{\rm CP}\hskip-1.80em{/}}}%temp replacement due to no font

\def\ap#1#2#3   {{\em Ann. Phys. (NY)} {\bf#1} (#2) #3.}
\def\apj#1#2#3  {{\em Astrophys. J.} {\bf#1} (#2) #3.}
\def\apjl#1#2#3 {{\em Astrophys. J. Lett.} {\bf#1} (#2) #3.}
\def\app#1#2#3  {{\em Acta. Phys. Pol.} {\bf#1} (#2) #3.}
\def\ar#1#2#3   {{\em Ann. Rev. Nucl. Part. Sci.} {\bf#1} (#2) #3.}
\def\cpc#1#2#3  {{\em Computer Phys. Comm.} {\bf#1} (#2) #3.}
\def\err#1#2#3  {{\it Erratum} {\bf#1} (#2) #3.}
\def\ib#1#2#3   {{\it ibid.} {\bf#1} (#2) #3.}
\def\jmp#1#2#3  {{\em J. Math. Phys.} {\bf#1} (#2) #3.}
\def\ijmp#1#2#3 {{\em Int. J. Mod. Phys.} {\bf#1} (#2) #3.}
\def\jetp#1#2#3 {{\em JETP Lett.} {\bf#1} (#2) #3.}
\def\jpg#1#2#3  {{\em J. Phys. G.} {\bf#1} (#2) #3.}
\def\mpl#1#2#3  {{\em Mod. Phys. Lett.} {\bf#1} (#2) #3.}
\def\nat#1#2#3  {{\em Nature (London)} {\bf#1} (#2) #3.}
\def\nc#1#2#3   {{\em Nuovo Cim.} {\bf#1} (#2) #3.}
\def\nim#1#2#3  {{\em Nucl. Instr. Meth.} {\bf#1} (#2) #3.}
\def\np#1#2#3   {{\em Nucl. Phys.} {\bf#1} (#2) #3.}
\def\pcps#1#2#3 {{\em Proc. Cam. Phil. Soc.} {\bf#1} (#2) #3.}
\def\pl#1#2#3   {{\em Phys. Lett.} {\bf#1} (#2) #3.}
\def\prep#1#2#3 {{\em Phys. Rep.} {\bf#1} (#2) #3.}
\def\prev#1#2#3 {{\em Phys. Rev.} {\bf#1} (#2) #3.}
\def\prl#1#2#3  {{\em Phys. Rev. Lett.} {\bf#1} (#2) #3.}
\def\prs#1#2#3  {{\em Proc. Roy. Soc.} {\bf#1} (#2) #3.}
\def\ptp#1#2#3  {{\em Prog. Th. Phys.} {\bf#1} (#2) #3.}
\def\ps#1#2#3   {{\em Physica Scripta} {\bf#1} (#2) #3.}
\def\rmp#1#2#3  {{\em Rev. Mod. Phys.} {\bf#1} (#2) #3.}
\def\rpp#1#2#3  {{\em Rep. Prog. Phys.} {\bf#1} (#2) #3.}
\def\sjnp#1#2#3 {{\em Sov. J. Nucl. Phys.} {\bf#1} (#2) #3.}
\def\spj#1#2#3  {{\em Sov. Phys. JEPT} {\bf#1} (#2) #3.}
\def\spu#1#2#3  {{\em Sov. Phys.-Usp.} {\bf#1} (#2) #3.}
\def\zp#1#2#3   {{\em Zeit. Phys.} {\bf#1} (#2) #3.}

\setcounter{secnumdepth}{2} % Number sections and subsections

%%%%%%%%%%%%%%%%%%%%%%%%%%%%%%%%%%%%%%%%%%%%%%%%%%
%                                                %
%    BEGINNING OF TEXT                           %
%                                                %
%%%%%%%%%%%%%%%%%%%%%%%%%%%%%%%%%%%%%%%%%%%%%%%%%%

\title{TEXTURES, FAMILY SYMMETRIES, AND NEUTRINO MASSES}

\firstauthors{ELENA PAPAGEORGIU}

\firstaddress{Laboratoire de Physique Th\'eorique et Hautes Energies,
Universit\'e de Paris XI, B\^atiment 211, 91405 Orsay, France}

%\secondauthors{ A.N. Other }

%if there are no second authors then comment out the line and adjust the
%maketitle command and \def\secondauthor... command in snow.sty

%\secondaddress{Department of Exotic Results, University of Heavens,
%Universe Road 999,\\  Whoknowswhere ZZZ123, Paradise}

%if there are no second authors then comment out the line and adjust the
%maketitle command and \def\secondaddress... command in snow.sty

\twocolumn[\maketitle\abstracts{
We review recent work on the interplay between broken family symmetries
which give rise to approximate texture zeros in the mass matrices of the quarks
and leptons and the presence of certain hierarchy patterns in the
spectrum of massive neutrinos.}]

\section{Introduction}
In an attempt to understand the hierarchy of the fermion masses and the mixing
of the quarks in terms of a few fundamental parameters the idea of extending
the Standard Model  (SM) by a ${\tilde U(1)}$ horizontal symmetry  of stringy
origin has been reinvestigated \cite{seiberg,ross,pap,bin,savoy,eap}
in the prospect of anomaly cancellation {\`a la Green Schwarz},
whereby, the mixed anomalies of the ${\tilde U(1)}$ with the $SU(3)$, the
$SU(2)$ and the $U(1)_Y$ gauge groups
are cancelled by a shift in the dilaton field
in 4D string theories, a sufficient condition
for obtaining also the canonical unification of the
three gauge couplings at $M_G = 10^{16}$ GeV,
as well as the phenomenologically successful
\begin{equation}
m_{\tau}\simeq m_b \qquad\qquad m_e \cdot m_{\mu}\simeq
m_d \cdot m_s
\end{equation}
relations without the need of a grand unified group.

Besides, since with the thirteen observables of the Yukawa sector one cannot
uniquely determine the entries of the quark and lepton mass matrices,
even by assuming that the latter are hermitian,
it has become increasingly popular to assume that some of the entries are
sufficiently suppressed with respect to others so that they can be replaced by
so-called ``texture zeros'' which reflect the nature of the underlying
horizontal symmetry and
give relations between the fermion masses and the mixings, like the successful
$|V_{us}| \simeq \sqrt{m_d/m_s}$ prediction.
We shall therefore start with a discussion of how to generate such
phenomenologically
successful symmetric textures
in the context of a broken horizontal symmetry with a minimum number of scalar
fields and higher-dimension operators.

The extension of this approach to the lepton
sector has given different neutrino mass
and oscillation patterns
\cite{eap,ep1,ep2,lola1,ramond,lola2} which are able to explain some
of the three neutrino-mass related puzzles, the solar neutrino deficit (SN),
the atmospheric neutrino deficit (AN), and the need of hot dark matter (HDM)
for galaxy formation. This will be discussed in the second part of the talk.

\section{Generating the texture of the quark and lepton mass matrices from an
extra ${\tilde U}(1)$}
Let us assume the existence of a family-dependent ${\tilde U}(1)$
symmetry  at the Planck scale, with respect to
which the quarks and leptons carry charges $\alpha_i$ and $a_i$
respectively, where $i=1,2,3$ is the generation index.
Due to the electroweak symmetry the lefthanded
up (quark/lepton) states and the corresponding down (quark/lepton)
states carry the same charges.
The generation of symmetric textures implies on the other hand  equal charges
for lefthanded and righthanded states.

We generate first the texture of the quark mass matrices $M_u$ and $M_d$.
Given the role played by the third
generation we start with rank-one matrices and make a choice for the charges
such  that only the (3,3) renormalizable couplings $t^c t h_1$ and $b^c b h_2$
are allowed, the other entries being zero as long as the symmetry is exact.
This choice fixes the charges of the light Higges $h_{1,2}$ to $-2\alpha$
($\alpha\equiv \alpha_3$). On the other hand the presence of scalar fields,
some of which acquire a vev at the unification scale,
and of higher-dimension operators, as this is common in most string
compactification schemes, can lead to the spontaneous breaking
of the symmetry and the generation of small
nonzero entries. The most economical scenario \cite{ross,pap} requires
only one singlet field or a pair of fields ${\tilde \sigma}_{\pm}$
developping equal (vev's) along a ``D-flat'' direction and carrying
opposite charges $\pm 1$.
These in turn can give rise to higher-order couplings
$q^c_{i} h_1 ({<{\tilde \sigma}>\over M})^{|2\alpha - \alpha_i - \alpha_j|}
q_{j}$
where M is a scale typical of these higher-dimension operators, {\it e.g.} the
string
unification scale $M_S \simeq 10^{18}$ GeV or $M_P$.
The power with which the scale ${\tilde {\cal E}} = {<{\tilde \sigma}>\over M}$
will fill in the
$(i,j)$ entry is such as to compensate the charge of $q^c_{i} h q_{j}$.
Notice that when the exponent is positive (negative) only the field
${\tilde \sigma}_+$ $({\tilde \sigma}_-)$ can contribute.
Most likely such a theory will contain also heavy Higgs multiplets $H_i$,
needed
for breaking the GUT symmetry and giving
rise to $q^c_{i} H ({<{\tilde \sigma}>\over M})^{|2\beta - \alpha_i -
\alpha_j|} q_{j}$,
where by $-2\beta_i$ we denote the charge of $H_i$.
As a consequence one can generate mass matrices of the following type:
\begin{equation}\label{Yz}
M_{u,d} \sim \left(
\begin{array}{ccc}
{\tilde {\cal E}}^{2 |z_1|} & {\tilde {\cal E}}^{|z_1 + z_2|}
& {\tilde {\cal E}}^{|z_1 + z|}\\
{\tilde {\cal E}}^{|z_1 + z_2|} & {\tilde {\cal E}}^{2 |z_2|}
& {\tilde {\cal E}}^{|z_2 + z|}\\
{\tilde {\cal E}}^{|z_1 + z|} & {\tilde {\cal E}}^{|z_2 + z|}
& 1 + {\tilde {\cal E}}^{2 |z|}\\
\end{array}
\right) \,,
\end{equation}
whith $|z_i| = |\beta - \alpha_i|$, and
$|z| = |\beta - \alpha|$.
Assuming two scales ${\tilde {\cal E}}\simeq \lambda\simeq 0.2 $
and ${\cal E} \simeq  {\tilde {\cal E}}^2$, where $\lambda\simeq |V_{us}|$ is
the {\it Wolfenstein} parameter, and the right combination of charges one can
generate four sets of $M_u - M_d$ textures: {\bf A}, {\bf B}, {\bf C} and {\bf
D}
(Table 1 and eq.(3)) which contain five
zeros in total \cite{eap}, without counting zeros of symmetric entries twice,
and lead to the most predictive {\it Ansatz} for quark masses and mixings in
agreement with experiment. A fifth set, {\bf E}, with four zeros emerges
from the particularly economical scenario with the two singlets
\cite{ross,pap}.
All five sets can be obtained from
\begin{equation}
{M_u \over m_t} \simeq \left(
\begin{array}{ccc}
0 & \alpha \lambda^6 & \delta \lambda^4 \\
\alpha \lambda^6 & \beta \lambda^4 & \gamma \lambda^2 \\
\delta \lambda^4 & \gamma \lambda^2 & 1
\end{array}
\right), \,
{M_d \over m_b} \simeq \left(
\begin{array}{rcl}
0 & {\tilde \alpha}\lambda^3 & 0 \\
{\tilde \alpha}\lambda^3 & {\tilde \beta}\lambda^2 & {\tilde\gamma}\lambda \\
0 & {\tilde \gamma}\lambda & 1
\end{array}
\right)
\end{equation}
and Table 1.

\begin{table}\begin{center}\caption{Possible sets of $M_u$, $M_d$ textures
corresponding to different choices of ${\tilde U}(1)$ charges and
classified according to eq.(3). The ones and the zeros correspond to an order
of magnitude estimate, since in our approach it is possible to generate
the powers of lambda but not the coefficients in front.}
\vspace{0.5cm}
\begin{tabular}{c|ccccccc} \hline\hline
& & & & & & &\\
{\bf Set} & $\alpha$ & $\beta$ & $\gamma$ & $\delta$ & ${\tilde \alpha}$ &
${\tilde \beta}$ & ${\tilde \gamma}$ \\ \hline
& & & & & & & \\
{\bf A} & 1 & 1 & 0 & 0 & 1 & 1 & 1\\
& & & & & & & \\
{\bf B} & 1 & 0 & 1 & 0 & 1 & 1 & 1\\
& & & & & & & \\
{\bf C} & 1 & 1 & 1 & 0 & 1 & 1 & 0\\
& & & & & & & \\
{\bf D} & 0 & 1 & 1 & 1 & 1 & 1 & 0\\
& & & & & & & \\
{\bf E} & 1 & 1 & 1 & 0 & 1 & 1 & 1\\
 \hline\hline
\end{tabular}
\end{center}
\end{table}

Let us now turn to the charged lepton mass matrix $M_e$.
Assuming simply the gauge symmetries of the SM the $U(1)_X$ charges of the
leptons  are not related to those of the quarks
but adopting the philosophy that all the entries except the (3,3) entry are
zero before symmetry breaking leads to $a_3 = \alpha$.
Another constraint comes from the second mass relation
of eq.(1) which implies $a_1 + a_2 = \alpha_1 + \alpha_2$.
This relation can be satisfied when $a_1 = \alpha_1$ and
$a_2 = \alpha_2$ and leads to the GUT relation:
\begin{equation}
M_e \sim M_d
\end{equation}
with a slight modification of the (2,2) entries as in the {\it Georgi-Jarlskog}
model.

Let us discuss next the generation of Dirac neutrino mass
terms: $M^D N_i^c \nu_j$ where $N_{i=1,2,3}$ are the three
righthanded neutrino states which are present in most GUT's.
Since these are of the same type as the
mass terms in the quark and charged lepton sector it is natural
to adopt again the same approach. Then, because the charges $a_i$ have been
fixed through the charged lepton {\it Ansatz},
$M^D = M_e$.
Furthermore for the choice $a_1 = \alpha_1$ and $a_2 = \alpha_2$ one obtains
the other well known GUT relation:
\begin{equation}
M^D = M_u \qquad {\rm or} \qquad M^D = M_d \,.
\end{equation}

On the other hand, the heavy Majorana mass terms
$M_R N_i^c N_j$, which are needed in the $(6\times 6)$ neutrino mass matrix
in order to suppress the otherwise unacceptably large masses for the
$\nu_e, \nu_{\mu}, \nu_{\tau}$, need not be
generated the same way. In compactified string models, due
to the absence of large Higgs representations, righthanded
neutrinos donot get tree-level masses, so all entries
in $M_R$ are due to nonrenormalizable operators \cite{Ranfone1}
$N_i^c H_k H_l N_j$ $(k,l=1,2,...)$ whose scale is of
 ${\cal O}(M_G^2/M_P)$ multiplied for some orbifold
suppression factor ${\cal C} \simeq 1-10^{-4}$:
\begin{equation}
R = {\cal C}\, {<H> <H>\over M_P}\, \simeq 10^9 - 10^{13} {\rm GeV} \,.
\end{equation}
Nothing is a priori known concerning the particular
texture of $M_R$ or the existence of hierarchy
in this sector.
One can therefore follow either of the two paths:
Assume no hierarchy and use the operators $N_i^c H_k H_l N_j$  to generate
a nonsingular Majorana mass texture
\footnote{otherwise some of the standard neutrinos will be too heavy and will
have to decay as required by various astrophysical and cosmological
considerations \cite{Ranfone2},}
\begin{equation}
M_R = \left(
\begin{array}{ccc}
R_1 & R_4 & R_5 \\
R_4 & R_2 & R_6 \\
R_5 & R_6 & R_3
\end{array}
\right) \,,
\end{equation}
as in refs.[3,6,8],
or generate the entries along the same principle which was used to generate the
hierarchy in the quark and charged lepton sector, refs.[9-11].
In any case the spectrum of the light neutrino states $\nu_e, \nu_{\mu},
\nu_{\tau}$ depends very delicately on the (2,3) and/or the (1,3) entry of
$M_{u,d}$ and the symmetries of $M_R$ \cite{eap,ep1,ep2}, as will be shown
next.

\section{The neutrino mass spectrum}
Starting from the {\it Ansatz}:
$ M_R \simeq {\bf 1}\times R$ or $M_R\sim M^{D}$
one is led, upon diagonalisation of the reduced
$(3\times 3)$ light neutrino matrix
\begin{equation}
M_{\nu}^{eff} \simeq  M^{D\dag} M_R^{-1} M^{D}
\,,
\end{equation}
to the quadratic seesaw spectrum:
\begin{equation}
m_{\nu_e} : m_{\nu_{\mu}} : m_{\nu_{\tau}} \simeq ( z^8 : z^4 : 1 )\times
m_{\nu_{\tau}}
\,,
\end{equation}
where
\begin{equation}
z=\lambda^2 \qquad m_ {\nu_{\tau}}= {m_t^2\over R} \quad {\rm if} \quad M^D =
M_u
\,,
\end{equation}
or,
\begin{equation}
z=\lambda \qquad  m_{\nu_{\tau}} = {m_b^2\over R} \quad {\rm if} \quad M^D =
M_d
\,,
\end{equation}
and the lepton mixings:
\begin{equation}
|V_{e - \mu}| \sim {\lambda \over 3}, \,
|V_{\mu-\tau}| \sim \lambda^2 - \lambda^3, \,
|V_{e-\tau}| \sim \lambda^4 - \lambda^5 \,,
\end{equation}
giving rise to neutrino oscillations with
\begin{equation}
sin^2 2\theta_{e-\mu} \sim 0.02, \,
sin^2 2\theta_{\mu-\tau} \sim 10^{-3}, \,
sin^2 2\theta_{e-\tau} \sim  10^{-5} \,.
\end{equation}
The eqs.(10-13) lead to an elegant resolution of the SN problem
through matter enhanced $\nu_e \to \nu_{\mu}$
oscillations \cite{petcov} while the  $\nu_{\tau}$ mass falls whithin
the range needed for it to be the HDM candidate.
The quadratic seesaw spectrum is typical of a much larger class of models when
$M_R$ does not contain any texture
zeros nor any symmetry \cite{ep1}. This becomes transparent when
$M_{\nu}^{eff}$
is expressed in terms of
the minors
of the matrix $M_R$, $r_{i = 1, ... 6}$, which
are obtained by omitting the row and column containing the corresponding $R_i$
entry, {\it e.g.},
$r_3 = R_1 R_2 - R_4^2$:
\begin{displaymath}
{M_{\nu}^{eff}\over m_{t,b}^2 /\Delta} =  r_3  \left(
\begin{array}{ccc}
{\delta^\prime}^2 z^4 & {\gamma^\prime} {\delta^\prime} z^3 & {\delta^\prime}
z^2 \nonumber\\
{\gamma^\prime} {\delta^\prime} z^3 & {\gamma^\prime}^2 z^2 & {\gamma^\prime} z
\nonumber\\
{\delta^\prime} z^2 & {\gamma^\prime} z & 1 \nonumber\\
\end{array}
\right) \qquad \nonumber
\end{displaymath}
\begin{displaymath}
\, + \,
r_6 z  \left(
\begin{array}{ccc}
 & ({\beta^\prime} {\delta^\prime} + {\alpha^\prime}{\gamma^\prime}) z^3  &
({\alpha^\prime}+ {\gamma^\prime} {\delta^\prime}) z^2 \\
({\beta^\prime} {\delta^\prime} + {\alpha^\prime}{\gamma^\prime}) z^3 & 2
{\beta^\prime}{\gamma^\prime} z^2 & ({\gamma^\prime}^2+{\beta^\prime}) z \\
 ({\alpha^\prime}+ {\gamma^\prime} {\delta^\prime}) z^2 &
({\gamma^\prime}^2+{\beta^\prime}) z & 2{\gamma^\prime}  \\
\end{array}
\right)
\end{displaymath}
\begin{displaymath}
\, + \,
r_2 z^2  \left(
\begin{array}{ccc}
{\alpha^\prime}^2 z^4 & {\alpha^\prime}{\beta^\prime} z^3 &
{\alpha^\prime}{\gamma^\prime} z^2 \\
{\alpha^\prime}{\beta^\prime} z^3 & {\beta^\prime}^2 z^2 &
{\beta^\prime}{\gamma^\prime} z \\
{\alpha^\prime}{\gamma^\prime} z^2 & {\beta^\prime}{\gamma^\prime} z &
{\gamma^\prime}^2 \\
\end{array}
\right)
\end{displaymath}
\begin{displaymath}
 \, + \,
r_5 z^2  \left(
\begin{array}{ccc}
 &  & {\delta^\prime}^2 z^2 \\
 & 2{\alpha^\prime}{\gamma^\prime} z^2 & ({\alpha^\prime}+ {\gamma^\prime}
{\delta^\prime}) z \\
{\delta^\prime}^2 z^2 & ({\alpha^\prime}+ {\gamma^\prime} {\delta^\prime}) z &
2 {\delta^\prime} \\
\end{array}
\right)
\end{displaymath}
\begin{displaymath}
\, + \,
r_4 z^3  \left(
\begin{array}{ccc}
 & {\alpha^\prime}^2 z^3 &  \\
{\alpha^\prime}^2 z^3 & 2{\alpha^\prime}{\beta^\prime} z^2 &
({\alpha^\prime}{\gamma^\prime} + {\beta^\prime}{\delta^\prime}) z \\
 & ({\alpha^\prime}{\gamma^\prime} + {\beta^\prime}{\delta^\prime}) z &
2{\gamma^\prime}{\delta^\prime} \\
\end{array}
\right)
\end{displaymath}
\begin{equation}
\qquad \, + \,
r_1 z^4  \left(
\begin{array}{ccc}
 &  &  \\
 & {\alpha^\prime}^2 z^2 & \\
 &  &  {\delta^\prime}^2 \\
\end{array}
\right)
\end{equation}
where $\Delta$ is the determinant of $M_R$, $z$ is as in eq.(10) or eq.(11),
and
$\left( {\alpha^\prime}, {\beta^\prime}, {\gamma^\prime}, {\delta^\prime}
\right)$ take the
values of $\left( \alpha,\beta,\gamma,\delta \right)$ if $M^D = M_u$, or, the
values of $\left( {\tilde\alpha}, {\tilde\beta}, {\tilde\gamma} \right)$ if
$M^D = M_d$,
as shown in Table 1.
When there is not much hierarchy among the $R_{i=1,...,6}$  the entries of
$M_{\nu}^{eff}$
follow the usual pattern:
$(3,3) > (2,3) > (2,2);(1,3)$, as long as $r_3 \not= 0$. This gives
rise to spectra which are only slightly distorted with respect to the quadratic
seesaw spectrum when due to the texture zeros and/or the symmetries of $M_R$
some of the other minors are zero.
In contrast when $r_3=0$, the hierarchy in the neutrino spectrum
can be reversed as this is the case for the quark texture solution
{\bf A} where a texture zero appears at the (3,3) entry of
$M_{\nu}^{eff}$.
Then, depending upon the position of the zeros in $M_R$, the muon neutrino
is heavier than the tau neutrino, or, they are mass degenerate and
$sin^2 2\theta_{\mu-\tau}\sim {\cal O}(1)$ \cite{ep2}.
Large $\nu_e-\nu_{\tau}$ mixing is on the other hand typical
of the quark texture solutions {\bf B} and {\bf C} \cite{ep2}.
As a matter of fact, mass-degenerate neutrinos and large mixing angles
emerge more naturally from the five-zero texture solutions
{\bf A}-{\bf D} \cite{eap,ep1,ep2}
than from  the four-zero solution {\bf E}
which requires very particular conditions of strong hierarchy in $M_R$
to give mass degenerate neutrinos  \cite{lola1,ramond,lola2}.
This may be so because the more texture zeros there are in the quark mass
matrices the more the symmetries of the righthanded neutrino sector can
influence the light neutrino spectrum.

\end{document}